\documentclass[reprint]{revtex4-1}

\usepackage{natbib}
\usepackage{graphicx}
\usepackage{amsmath,amssymb}
\usepackage{txfonts}
\usepackage{epstopdf}
\usepackage{natbib}
\usepackage{changes}
\definechangesauthor[color=red]{RV}
\definechangesauthor[color=blue]{LM}
\definechangesauthor[color=olive]{PH}
\definechangesauthor[color=violet]{SL2}
\definechangesauthor[color=violet]{SL}

\newcommand{\vect}[1]{{\boldsymbol{#1}}}

\begin{document}
\title{Spectral anisotropy and intermittency of plasma turbulence at ion kinetic scales}
\author{Simone~Landi}
\affiliation{Dipartimento di Fisica e Astronomia, Universit\`a di Firenze,
Firenze, Italy.}
\affiliation{Osservatorio Astrofisico di Arcetri, INAF, Firenze, Italy}
\author{Luca~Franci}
\affiliation{School of Physics and Astronomy, Queen Mary University of London, London, UK}
\author{Emanuele~Papini}
\affiliation{Dipartimento di Fisica e Astronomia, Universit\`a di Firenze,
Firenze, Italy.}
\author{Andrea~Verdini}
\affiliation{Dipartimento di Fisica e Astronomia, Universit\`a di Firenze,
Firenze, Italy.}
\author{Lorenzo~Matteini}
\affiliation{LESIA, Observatoire de Paris, Universit\'e PSL, CNRS, Sorbonne Universit\'e, Univ. Paris-Diderot, Sorbonne Paris Cit\'e\\place J. Janssen 5, F-92192 Meudon, France}
\altaffiliation{Department of Physics, Imperial College London, London SW7 2AZ, UK.}
\author{Petr~Hellinger}
\affiliation{Astronomical Institute, CAS, Bocni II/1401,CZ-14100 Prague, Czech Republic}
\affiliation{Institute of Atmospheric Physics, CAS, Bocni II/1401, CZ-14100 Prague, Czech Republic}

\date{\today}
 
\begin{abstract}
By means of three dimensional high-resolution hybrid simulations we study the properties of the magnetic field spectral anisotropynear and beyond ion kinetic scales. By using both a Fourier analysis and a local analysis based on multi-point 2nd-order structure function techniques, we show that the observed anisotropy is less than what expected by standard wave normal modes turbulence theories although the non linear energy transfer is still in the perpendicular direction with respect to the magnetic field, only advected in the parallel direction as expected balancing the non-linear energy transfer time and the decorrelation time. Such result can be explained by a phenomenological model based on the formation of strong intermittent two-dimensional  structures in the plane perpendicular to the local mean field that fullfill some prescribed aspect ratio eventually depending on the scale.  This model supports the idea that small scales structures, such as reconnecting current sheets, contribute significantly to the formation of the turbulent cascade at kinetic scales.
\end{abstract}

%\pacs{47.65.-d, 47.27.Jv, 47.27.Gs}
\maketitle

%**********************************************************
\section{Introduction}
The solar wind plasma is coupled to electromagnetic fluctuations whose spectra follow power laws spanning a huge range of frequencies and are the signature of a strong turbulent activity that taking place in there \citep[e.g.][]{Bruno_Carbone_2013,Kiyani_al_2015,Chen_2016}. 
%\deleted[id=SL2]{In-situ measurements of solar wind plasma and electromagnetic fields show spectra with power-laws scaling spanning several decades in frequency}, $f$ \citep[e.g.][]{Bruno_Carbone_2013,Kiyani_al_2015,Chen_2016}.  \deleted[id=SL2]{Power-laws support an interpretation in terms of turbulent fluctuations, although the rich variety of spectral features is not easily explained in the framework of known turbulent theories and phenomenologies.}
At $1~\mathrm{AU}$, plasma fluctuations with frequencies in the interval $10^{-4}~\mathrm{Hz}\lesssim f\lesssim 10^{-2}~\mathrm{Hz}$ are reasonably well described by the one-fluid magneto-hydrodynamic (MHD) equations: there, the spectra of the magnetic field, the ion bulk velocity, and the electric field are dominated by the transverse components with respect to the ambient magnetic field $\vect{B}_0$.  Magnetic fluctuations typically show a Kolmogorov-like spectrum (i.e., having the -5/3 slope), while the velocity field power is slightly shallower (typically close to -3/2) \citep[e.~g.][]{Podesta_al_2007,Salem_al_2009,Roberts_2010,Wicks_al_2011,Tessein_al_2009} and strongly coupled which electric field fluctuations  \citep{Chen_al_2011b} as expected in the ideal MHD regime. This regime is the so-called MHD inertial range.

%\deleted[id=SL2]{For frequencies corresponding to the magneto-hydrodynamic (MHD) range, $10^{-4}~\mathrm{Hz}\lesssim f\lesssim 10^{-2}~\mathrm{Hz}$ at $1~\mathrm{AU}$, spectra of the magnetic field, the ion bulk velocity, and the electric field are dominated by the transverse components with respect to the ambient magnetic field $\vect{B}_0$.  Magnetic fluctuations typically show a Kolmogorov-like spectrum (i.e., having the -5/3 slope), while the velocity field power is slightly shallower (typically close to -3/2)} \citep[e.~g.][]{Podesta_al_2007,Salem_al_2009,Roberts_2010,Wicks_al_2011,Tessein_al_2009} \deleted[id=SL2]{and strongly coupled which electric field fluctuations  \citep{Chen_al_2011b} as expected in the ideal MHD regime.} 

At higher frequencies, $f\gtrsim 10^{-1}~\mathrm{Hz}$, corresponding to length scales approaching typical ion kinetic scales, there is a change in the properties of the observed fluctuations. Near these scales, a spectral transition (break) appears in the magnetic field fluctuations, separating the MHD inertial range from a second, steeper, power-law interval at kinetic scales. There is a certain degree of variability of the spectral index just below this transition \citep[e.~g.][]{Leamon_al_1998, Smith_al_2006b,Sahraoui_al_2010}, although the slope seems more universal at higher frequencies  with an index of $-2.8$ \citep{Alexandrova_al_2009,Alexandrova_al_2012}.
%The spectral index of magnetic fluctuations after the break varies between $(-4,-2)$ \citep[e.~g.][]{Leamon_al_1998, Smith_al_2006b,Sahraoui_al_2010}, although it tends to cluster around a slope of $-2.8$ for higher frequencies \citep{Alexandrova_al_2009,Alexandrova_al_2012}. 
At those scales, the ion bulk velocity decouples from magnetic field fluctuations and its spectrum shows (if any) an even steeper power-law slope \citep[e.g.][]{Safrankova_al_2016}; on the contrary, the electric field spectrum is shallower at sub-ion scales \citep{Bale_al_2005,Kellogg_al_2006} and becomes dominant over the magnetic field power, with a typical spectral index around $-0.8$ \citep[e.g.][]{Stawarz_al_2016, Matteini_al_2017}, consistent with the scaling predicted by the generalized Ohm's law \citep{Franci_al_2015b}. Which ion characteristic kinetic scale is the most relevant in determining the position of such transition has not been clearly identified, although observational evidences \citep{Chen_al_2014b,Bruno_Trenchi_2014} and numerical simulations \citep{Franci_al_2016} suggest that it is likely related to the larger between the ion inertial length and the ion gyroradius for extreme values of the the plasma beta, and to a combination of both in the intermediate-beta regime. At ion kinetic scales, also the nature of the fluctuations changes: an increase of the magnetic compressibility \citep{Salem_al_2012,Kiyani_al_2013} and a reduced variance anisotropy \citep{Podesta_TenBarge_2012} are observed; the density, while it follows a Kolmogorov spectrum at MHD scales, shows a flattening near the break  \citep{Safrankova_al_2015}, then becomes strongly coupled to the magnetic field fluctuations  at sub-ion scales\citep{Chen_al_2013}. 

%Numerical simulations in 2D are able to reproduce several aspects of the plasma turbulence either using a fluid  \citep[e.~g.][]{Shebalin_al_1983,Biskamp_Welter_1989,Ghosh_al_1996,Servidio_al_2009}  or kinetic \citep[e.~g.][]{Biskamp_al_1996,Biskamp_al_1999,Gary_al_2008,Parashar_al_2009,Markovskii_Vasquez_2010,Camporeale_Burgess_2011,Servidio_al_2012,Wan_al_2012,Franci_al_2015b} description. In particular \citet{Franci_al_2015a} and \citet{Franci_al_2015b} was able to reproduce turbulent spectra, with well defined power laws for several fields,  crossing the transition between fluid and ion scales and covering almost two decades on wave-number.  However, since turbulence is an inherited three-dimensional process, a more completed comprehension of the cascade can be achieved mainly by means of 3D numerical simulations. The polarizations properties of the turbulent fluctuations around the ion scales has been investigated in 3D simulations by \citet{Cerri_al_2017b} while a detailed analysis of the spectral properties of the main fields covering almost two decades in k-vectors  in the transition region has been reported by \citet{Franci_al_2018} while more recently a study of the dissipative properties have been done by \

An important aspect of plasma turbulence is how the cascade toward small scales proceeds with respect to the parallel and perpendicular component of a large scale mean magnetic field. At fluid scales, three-dimensional (3D) numerical simulations \citep[e.~g.][]{Oughton_al_1994,Cho_Vishniac_2000,Maron_Goldreich_2001,Cho_Lazarian_2002,Muller_Grappin_2005,Mason_al_2008, Beresnyak_Lazarian_2009a, Grappin_Muller_2010, Lee_al_2010, Perez_Boldyrev_2009,Boldyrev_al_2011, Chen_al_2011a}  have highlighted the role of spectral anisotropy in the formation of the MHD turbulent spectra, including the case when the radial expansion of the solar wind is taken into account \citep{Dong_al_2014,Verdini_Grappin_2015,Verdini_Grappin_2016}. In particular several numerical simulations \citep[e.g.][]{Maron_Goldreich_2001,Cho_Lazarian_2002,Verdini_al_2015} have shown that spectra tend to organize themselves in the so called critical balance state \citep{Goldreich_Sridhar_1995,Goldreich_Sridhar_1997,Galtier_al_2000}, in which the cascade proceeds by transferring energy only to scales where the non-linear interactions of eddies are faster than the linear propagating time of the normal modes of the fluid system. At such scales, this balance predicts that energy is transferred mostly in the direction perpendicular to the (local) mean field following a Kolmogorov cascade, while the parallel  spectrum should decrease as $k_\parallel^{-2}$. As a consequence, the two-dimensional (2D) spectral anisotropy should scale as  $k_\parallel \propto k_\perp^{2/3}$ \citep{Cho_Vishniac_2000,Cho_al_2002}. Observations seem to confirm to some extent that such energy transfer occurs in the solar wind plasma \citep{Bieber_al_1996,Horbury_al_2008,Podesta_2009,Wicks_al_2011,Verdini_al_2018a,Verdini_al_2019}.

The critical balance argument can be extended to the kinetic (sub-ion) scales, provided that one uses the electron-velocity eddy turn-over time as the relevant non-linear time and takes into account the dispersive nature of the fluctuations in this regime. Regardless of the normal mode used, either low-frequency strongly-oblique propagating  Kinetic Alfv\'en Waves (KAW) \citep{Schekochihin_al_2009} or quasi-parallel high-frequency whistler modes \citep{Biskamp_al_1996}, standard wave-wave interaction models of turbulence predict for the magnetic field spectrum a $k_{\perp}^{-7/3}$ scaling at sub-ion scales, and a further increase of the spectral anisotropy  $k_{\parallel}\propto k_{\perp}^{1/3}$ with a steeper (than MHD scales) parallel spectrum $\propto k_{\parallel}^{-5}$. A magnetic power decreasing as $k^{-7/3}$ has indeed been observed in several simulations using electron MHD  \citep[e. g.][]{Biskamp_al_1996,Biskamp_al_1999,Ng_al_2003,Cho_Lazarian_2004,Shaikh_2009}, Hall MHD \citep{Shaikh_Shukla_2009} and girokinetic models \citep{Howes_al_2008a}. Numerical evidence of the critically balanced scaling $k_\parallel\propto k_\perp^{1/3}$ have been reported by \citet{Cho_Lazarian_2004}. 

Steeper magnetic field spectra, $\propto k_{\perp}^{-2.8}$, in good agreement with solar wind observations, have been reproduced in girokinetic  simulations \citep{Howes_al_2011} where the steepening (with respect to the -7/3 prediction) was attributed to electron Landau damping. Using a frequency analysis \citet{TenBarge_Howes_2012} found that these simulations are in agreement with a $k_{\parallel}\propto k_{\perp}^{1/3}$ anisotropy scaling, although a perpendicular spectrum steeper than $k_\perp^{-7/3}$ should lead to a significant stronger anisotropy. 
Indeed, the critical balance condition both at MHD and kinetic scales can be written as $k_{\parallel}\propto k_{\perp} b(k_\perp)$, with $b(k_\perp)$ being the magnetic fluctuations amplitude at the scale $k_\perp$: as the spectrum becomes steeper, $b(k_\perp)$ steepens and the anisotropy increases. For a spectral slope of the magnetic power $P_B=b^2(k_\perp)/k_\perp\propto k_\perp^{-2.8}$, $b(k_\perp)\propto k_\perp^{-0.9}$ and the critical balance predicts $k_\parallel\propto k_\perp^{0.1}$. In this Kolmogorov-like phenomenological model only a slope $-7/3$ is consistent with the scaling $k_{\parallel}\propto k_{\perp}^{1/3}$.

Retaining the effects of linear ion and electron Landau damping, theoretical models \citep{Passot_Sulem_2015} and numerical simulations \citep{Passot_al_2014,Sulem_al_2016} predict a strong variability of the slope in the perpendicular spectrum together with an increase in the spectral anisotropy in the sub-ion range. Such variability is correlated with the ratio between the non-linear energy transfer time and the propagation/damping properties of low-frequency highly-oblique electromagnetic waves (the Kinetic Alfv\'en waves).  Some variability of the magnetic energy slope has been reported also in 2D \citep{Camporeale_Burgess_2011} and 3D full pic simulations \citep{Gary_al_2012}. 
However \citet{Hellinger_al_2018}, by analysing the properties of the 3rd order structure functions in 2D high resolution numerical simulations at low and moderate values of the plasma $\beta$, have shown that the sub-ion scales electromagnetic field power can be reasonably well described by an inertial (i.~e. dissipativeless) Hall-MHD range. Moreover, fluid 2D Hall-MHD simulations, where kinetic damping  effects are not taken into account, has been able to produce spectra steeper than $-7/3$ in remarkably good agreement with analogous hybrid kinetic simulations and  observations \citep{Papini_al_2019a}.

Several 3D numerical simulations investigating sub-ion scales have highlighted the role of coherent structures and intermittency in characterizing the turbulent cascade and its dissipation \citep[e. g.][]{Dmitruk_Matthaeus_2006a,Servidio_al_2015,Wan_al_2015}. \citet{Boldyrev_Perez_2012} have shown that, if energy is concentrated in spatially localized structures at small scales, the spectrum of magnetic fluctuations steepens following a $-8/3$ power law \citep[see also][]{Boldyrev_al_2013}, the anisotropy should follow $k_\parallel \propto k_\perp^{2/3}$, and the parallel spectrum a $-7/2$ power-law. \citet{Meyrand_Galtier_2013}, in the framework of EMHD, reported a similar perpendicular scaling but a steeper parallel spectrum, $\propto k_{\parallel}^{-5}$, consistent with a turbulent cascade driven merely by a 2D dynamics. More recently, \citep{Cerri_al_2017b} reported 3D hybrid Vlasov simulations where parallel and perpendicular spectra are qualitatively consistent with the model proposed by \citep{Boldyrev_Perez_2012} at $\beta_{\rm p}=1$ while steeper slopes are observed for a lower value of the proton plasma beta ($\beta_{\rm p}=0.2$). Concurrently, \citet{Franci_al_2018}, by analyzing the spectral properties of a high-resolution hybrid PIC simulation able to cover almost two decades in $k$-vectors, found that the spectral anisotropy, though large in the inertial range, becomes frozen in the sub-ion range following a  $k_\parallel \propto k_\perp$ scaling. A similar observation have been reported by \citet{Arzamasskiy_al_2019}. 
%Finally anisotropy consistent with the standard critical balance scenario, $k_\parallel\propto k_\perp^{1/3}$ has been reported by \citet{Groselj_al_2018} although 

In this work we present results from a high-resolution Particle-in-cell (PIC) hybrid (fluid electrons, kinetics protons) 3D numerical simulation of freely-decaying turbulence in presence of a mean magnetic field.  This simulation is very similar to that described in \citet{Franci_al_2018}, but uses a very high spatial grid resolution to focus on the spectral anisotropies at sub-ion scales. Fourier spectra show a transition near the ion scales and the simulation is able to produce a well defined spectrum in the kinetic range over more than one decade.  
A closer inspection of the magnetic and the density spectra shows that the spectral anisotropy with respect to the mean field stops to increase near the ion scales but the turbulent cascade still proceeds essentially in the perpendicular direction. A local analysis, using the multi-point 2nd-order structure function technique confirms that the anisotropy is frozen also with respect to the local mean field. Such result can be interpreted by generalizing the model proposed by \citep{Boldyrev_Perez_2012} and assuming that, at ion scales, the energy is mostly contained in intermittent coherent structures whose characteristic filling factor follows a prescribed scaling, . 
This seems to be consistent with models \citep{Mallet_al_2017b,Loureiro_Boldyrev_2017a,Boldyrev_Loureiro_2017,Loureiro_Boldyrev_2017b} and numerical simulations \citep{Franci_al_2017,Cerri_Califano_2017} which show that current sheets actively participate in the formation of the turbulent cascade.

%%%%%%%%%% NUME SETUP %%%%%%%%%%%%%

\section{Numerical setup}
We employ the hybrid particle-in-cell (HPIC) code CAMELIA (Current Advance Method Et cycLIc leApfrog) \citep{Franci_al_2018b}, where the electrons are considered as a massless, charge neutralizing fluid, whereas the ions (protons) are described by a particle-in-cell model (see \citet{Matthews_1994} for detailed model equations) and are advanced by the Boris scheme. Units of mass, length, and time are the ion (proton) mass $m_{\rm i}$, its inertial length $d_{\rm i}$ and the inverse of its gyroradius  $1/\Omega_{\rm i}$. We use $512^3$ collocation points in a spatial cubic grid with resolution $\Delta x=\Delta y=\Delta z= 0.0625 \, d_{\rm i}$ and $2048$ particles per cell (ppc) representing ions. Accumulation of energy at small scales is prevented by adopting a resistivity coefficient $\eta=1.5 \times 10^{-3}~4\pi v_Ac^{-1}\Omega_{\rm i}^{-1}$. The ions are advanced with a time step  $\Delta t=0.00625 \, \Omega_{\rm   i}^{-1}$, while the magnetic field $\boldsymbol{B}$ is advanced with a smaller time step $\Delta t_B = \Delta t/10$.
The initial condition  consists of a uniform and neutral plasma density $n=n_{\rm i}=n_{\rm e}$, a uniform magnetic field  directed along $z$, $\vect{B_0}=B_0\hat{\vect{z}}$, where species have isotropic and equal temperatures, $T_{\rm i} = T_{\rm e}$. The relative strength of the thermal and field pressure is measured in term of the ion (electron) plasma beta, $\beta_{\rm i,e}=8\pi n K_{\rm B} T_{\rm i,e} / B^2_0$, $K_{\rm B}$ being the Boltzmann's constant.
The setup of the simulation is similar to that shown in \citep{Franci_al_2018}: Decorrelated, Alfv\'enic-like fluctuations, initially isotropic in the 3D $k$-space vector range $0.20 < kd_{\rm i} < 0.80$, perturb the initial condition and are left to decay. The energy is equally partitioned between velocity and magnetic field with $B^{\rm rms}/B_0\sim 0.38$, the electron and proton temperatures are the same $\beta_{\rm e}=\beta_{\rm i}=0.5$.
The analysis here reported has been performed at the maximum of the turbulent activity, $t = t_{\rm max} \equiv 36~\Omega_{\rm i}^{-1}$, when the current density rms reaches its peak \citep{Franci_al_2018}. The maximum of turbulent activity time here is less than the simulation shown in \citet{Franci_al_2018}, simply because the energy is initially injected at smaller scales. Anyway, rms and physical quantities evolve in a very similar manner. 

%--------------- REDUCED 1D SPECTRA and COMPENSATED ISO 1D
\begin{figure}[t]
\begin{center}
\includegraphics[width=0.9\linewidth]{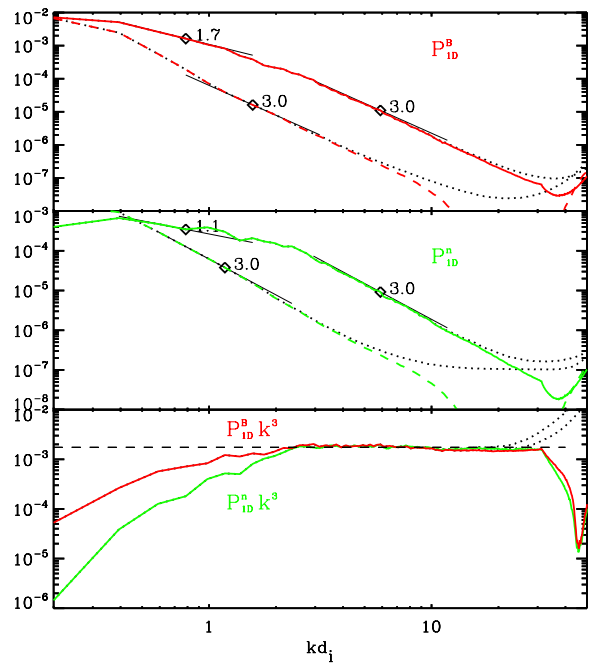}
\caption{1D reduced perpendicular (solid) and parallel (dashed) spectra of magnetic (top panel) and density fluctuations (middle). Power laws are drawn in thin solid black lines as a reference. On the bottom the isotropized 1D spectra of $B$ (red) and density $n$ (green) compensated by $k^3$. The spectra are obtained by applying a post-processing filtering procedure (see text for details). The unfiltered spectra are shown as dotted black curves. }
\label{fig:1DReducedSpectra}
\end{center}
\end{figure}
\section{Results}
 An overview of the spectral properties of the simulation is shown in Fig.~(\ref{fig:1DReducedSpectra}) where we report the magnetic (top) and density (midlle) omnidirectional (i. e., integrating the 3D power spectrum over all the  possible directions) power spectra as a function of the perpendicular (solid) and parallel (dashed) $k$-vectors
\begin{equation}
P_{\rm 1D}(k)\!\!=\!\!\!\!\int_0^{2\pi}\!\!\!\!\int_{-1}^1\!\!\!\!k^2 P(k\!\cos\!\phi\cos\!\theta,k\!\sin\!\phi\cos\!\theta,k\!\sin\!\theta) d\phi d(\cos\theta)~{.} 
\end{equation}
The original spectra, shown as dotted lines in each panel, are obtained by removing the numerical noise accumulated at small scales, in a way similar to that shown in \citep{Franci_al_2018}: we set to zero the amplitude of each 3D Fourier mode whose wave-number modulus is larger than a defined cut-off $k_{\rm cf}$. Moreover, we set to zero also all the modes of all the fields for which the magnetic energy density is less than a given noise threshold. Here the noise level has been choses as $10^{-11}$ and the cut-off is $k_{\rm cf}=2k_{\rm max}/3\simeq 170 k_0$ with $k_0=2\pi/L_x$ and $k_{\rm max}$ the Nyquist wave-vector.    
For both fields, the reduced spectra in the perpendicular direction show a well defined $-3$ power law over almost one-decade for $kd_{\rm i} >1$.  At large scales the filtered and unfiltered spectra are superimposed while the filtering procedure allows to extend the power-law in the high $k$-vector region. As expected, for both fields, at all scales the parallel reduced spectrum is subdominant with respect to its perpendicular counterpart, meaning that a spectral anisotropy is developed; however its power-law slope is the same as the perpendicular reduced spectrum, meaning that the spectral anisotropy produced at large scales $k_\perp d_{\rm i} < 1 $ is then frozen at sub-ion scales $k_\perp d_{\rm i} >1$.

%---------------------SPETTRO 2D--------------------
\begin{figure}[t]
\begin{center}
\includegraphics[width=1.1\linewidth]{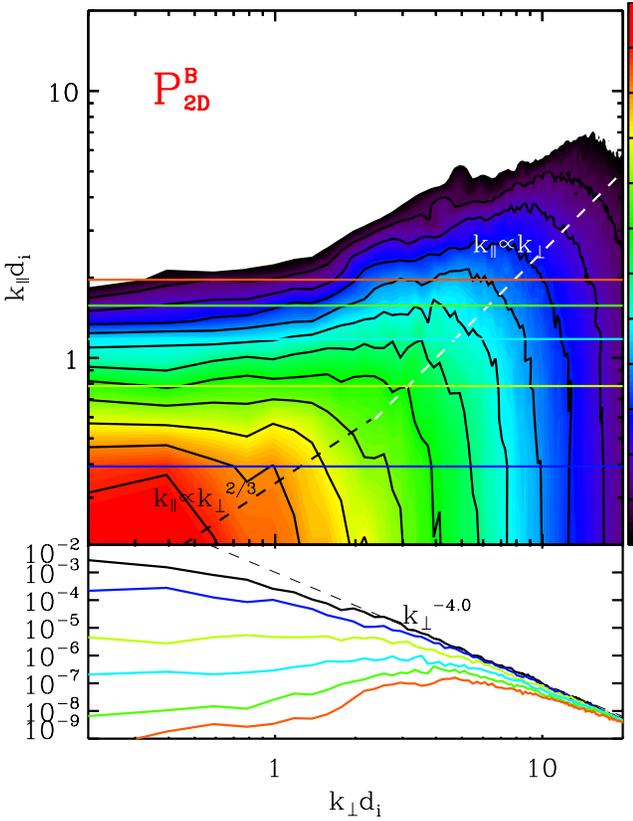}
\caption{Top panel: Reduced 2D spectrum of the total magnetic field fluctuations in  the $(k_\perp,k_\parallel)$ space (log-log scale) at $t=36 \Omega_{\rm i}^{-1}$.The dashed white line highlights the $k_\parallel \propto k_\perp$ scaling in the $2\le k_\perp\le 20$ region. A $k_\parallel \propto k_\perp^{2/3}$ (dashed black line) in the low $k_\perp$ domain is also included as reference. Bottom panel: Cuts of the 2D spectrum for $k_\parallel={\rm const}$: each color corresponds to a cut in $k_\parallel$ outlined with the same color in the top panel (the black line corresponds to $k_\parallel=0$).} 
\label{fig:2Dspectra}
\end{center}
\end{figure}
%------------------------------------------------------------------
% Funzioni di struttura locale
%------------------------------------------------------------------
\begin{figure*}[t]
\begin{center}
\includegraphics[width=0.44\linewidth]{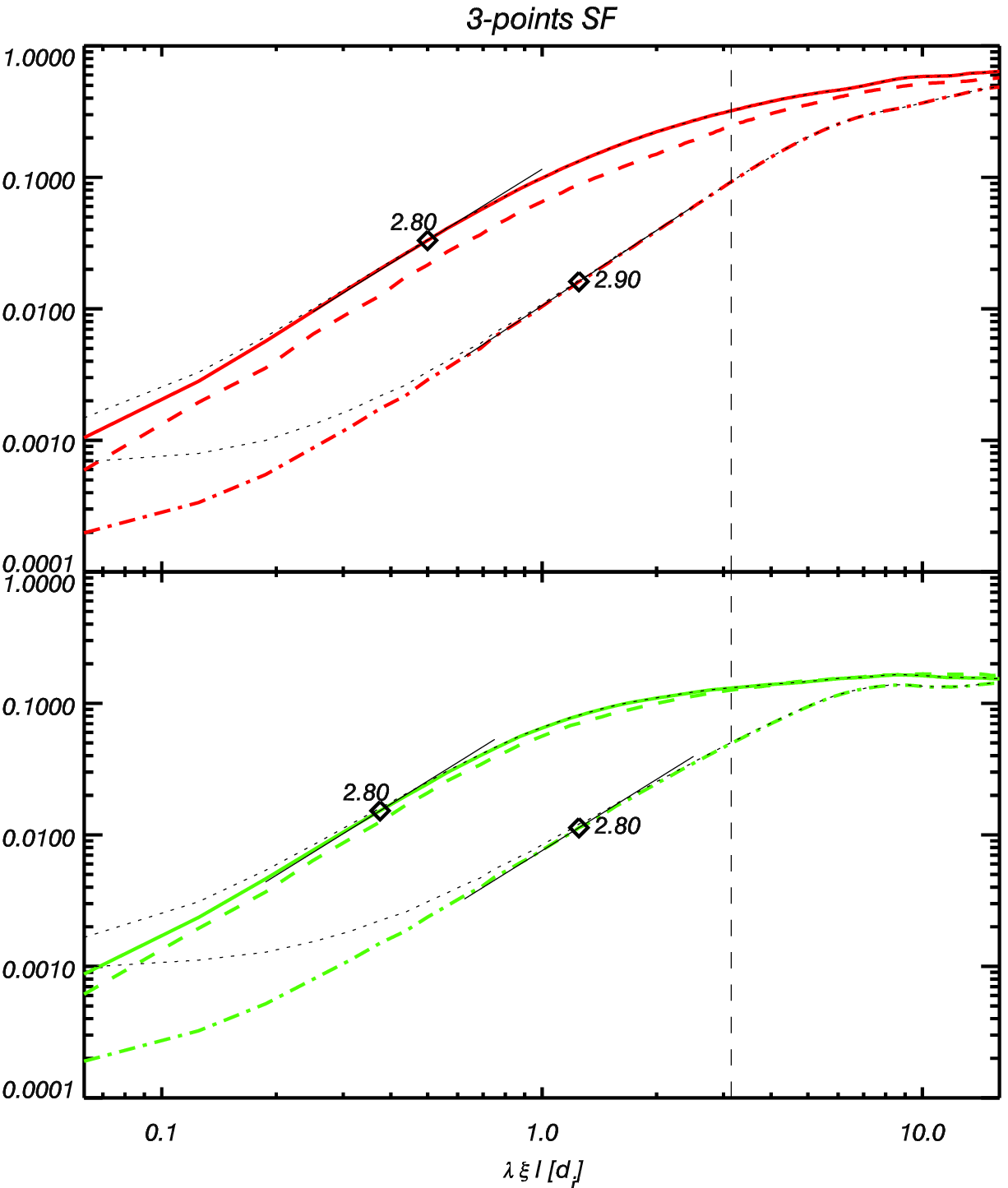}
\includegraphics[width=0.44\linewidth]{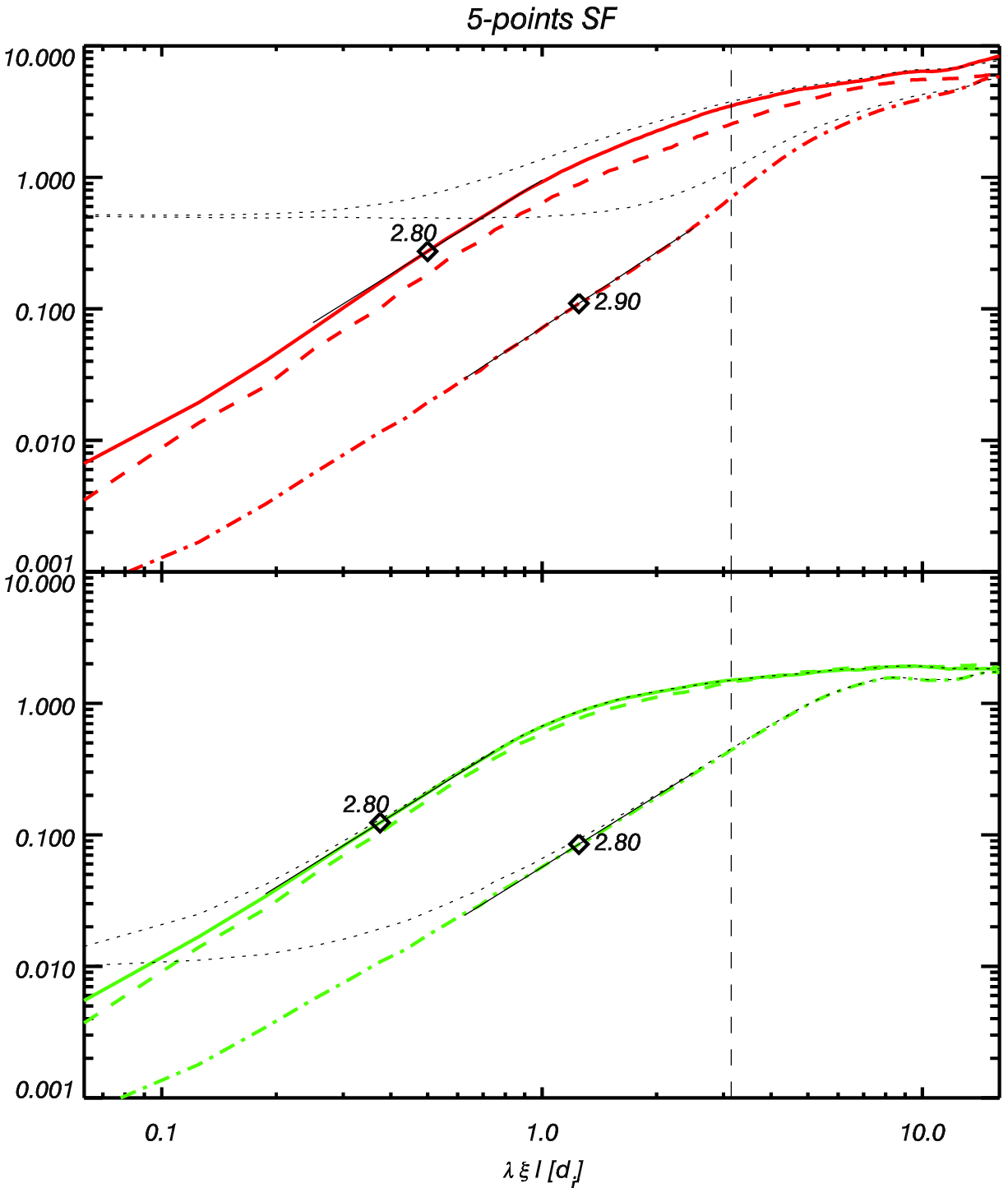}
\caption{Local 2nd-order 3 (left) and 5 (right) points structure functions for the magnetic (top), and density (bottom) fields. Solid and dashed lines are the two perpendicular components $\hat{\lambda}$ and $\hat{\xi}$ while the dash-dotted is for the parallel one $\hat{l}$. Power laws are drawn in thin solid black lines as a reference and are expressed in terms of the corresponding slopes in Fourier spectra. Dotted lines are the same quantities computed without using the filtering procedure outlined in the text.}
\label{fig:3PSF}
\end{center}
\end{figure*}
To understand the reduced 1D power spectra it is useful to analyse the reduced 2D spectra obtained by integrating the 3D spectra over all possible directions in the plane perpendicular to the main field.
\begin{equation}
P_{\rm 2D}(k_\parallel,k_\perp)=\int_{0}^{2\pi} k_\perp P(k_\perp\cos\phi,k_\perp\sin\phi,k_\parallel) d\phi~{.} 
\end{equation}
In Fig.~\ref{fig:2Dspectra}, the 2D reduced spectrum of the magnetic fluctuations in log-log scale is shown. At large scales  a spectral anisotropy is developed with the power mostly confined in the region $k_\perp > k_\parallel$. At higher wave-vectors,  $k_\perp d_{\rm i} \gtrsim 1$, this anisotropy appears to remain almost constant and most of the power seems confined in a region $k_\parallel \propto k_\perp$ (highlighted with the dashed white-line here used only as reference). Such a behaviour is different from the spectral anisotropy   predicted by the critical balance conditions in KAW or Whistler mediated turbulence ($k_\parallel \propto k_\perp^{1/3}$) \citep{Cho_Lazarian_2004,Schekochihin_al_2009}  and also to that proposed by \citep{Boldyrev_Perez_2012} where a spectral anisotropy $k_\parallel \propto k_\perp^{2/3}$ was obtained taking also into account the effect of the intermittency.   
A closer inspection to the 2D spectrum shows that in the region where the spectral anisotropy is frozen (constant aspect ratio) the iso-levels are almost independent of $k_\parallel$ meaning that the energy transfer occurs essentially in the perpendicular direction.
To better elucidate this, cuts with $k_\parallel={\rm const}$ of the 2D reduced spectra are shown in the bottom panel of Fig.~\ref{fig:2Dspectra}. For small value of $k_\parallel$, they have a significant level of energy at small value of $k_\perp$ and with different slopes. On the other hand, at approximately $k_\perp d_{\rm i}=2$ they all collapse and follow the same power-law which is well approximated by  $P^{\rm \mathbf{B}}_{\rm 2D}(k_\bot,k_{||}={\rm const})\propto k_\bot^{-4}$. Cuts with larger values of $k_\parallel$ are almost depleted of energy for small values of $k_\perp$ which then increases near $k_\perp={\rm C}k_\parallel$, following approximatively the same power law, $\propto k_\perp^{-4}$ (until small-scale numerical effects become dominant). 
 
At the lowest level of approximation, the 2D magnetic power spectrum at small scales can be described as a spectrum that is almost independent on $k_\parallel$ and bounded by the condition  $k_\perp\le{\rm C}k_\parallel^\alpha$ and where each perpendicular spectrum at constant $k_\parallel$ follows a  $k_\perp^{-\beta}$ power-law:
\begin{equation}
\left\{\begin{array}{cc}
P^{\rm \mathbf{B}}_{\rm 2D}=P_0 k_\perp^{-\beta}~&~{\rm if}~~ |k_\parallel |<k_\perp^{\alpha} \\
\\
0 & ~ {\rm Otherwise}
\end{array}
\right.
\end{equation}
Such simplified 2D model produces a reduced perpendicular spectrum $k_\perp^{\alpha-\beta}$, while the parallel one goes as $k_\parallel^{(1-\beta)/\alpha}$. If we now assume $\alpha\!=\!1$ and $\beta\!=\!4$ we obtain the same power-law  $-3$ for both perpendicular and parallel spectra. Within this model one should observe $\beta\!=\!10/3$  if the spectral anisotropy was the one proposed by \citep{Boldyrev_Perez_2012} ($\alpha\!=\!2/3$) while $\beta\!=\!6/3$ is the power-law required to obtained the reduced spectra in the self-similar model of turbulence ($\alpha\!=\!1/3$). Both values are significantly shallower than the behavior observed in our simulation.

One important question that arises here is whether the $z$-direction maps correctly the parallel direction of the (local) mean field ,that is, the relevant framework with respect to which the spectral anisotropy properties are supposed to be valid \citep{Cho_Vishniac_2000,Cho_al_2002,Cho_Lazarian_2004,Horbury_al_2008,Cho_Lazarian_2009,Wicks_al_2011}. 
To this aim, the scale-dependent anisotropy can be measured using the 2nd order structure function ($S\!F_2$) calculated in the local frame. 
However, since the slope of the fluctuations is near $-3$ in the sub-ion range, the usual $2$-points 2nd-order structure function is not suitable to study the spectral anisotropy at such scales and multi-points 2nd-order structure functions must be used \citep{Falcon_al_2007,Cho_Lazarian_2009}. 
Here we use the 3- and 5-points for a quantity $A$, defined respectively as 
\begin{equation}
S\!F_2^{(3)}({\bf r};A))\!=\!\langle\delta A^2\rangle\!=\!\langle(A({\bf x}\!+\!{\bf r})\!-\!2A({\bf x})\!+\!A({\bf x}\!-\!{\bf r}))^2\rangle
\end{equation}
and 
\begin{eqnarray}
&&S\!F_2^{(5)}({\bf r};A)\!=\!\langle\delta A^2\rangle  \nonumber \\
\! &&=\!\langle(A({\bf x}\!\!+\!\!2{\bf r})\!-\!4A({\bf x}\!\!+\!\!{\bf r})\!+\!6A({\bf x})\!-\!4A({\bf x}\!\!-\!\!{\bf r})\!+\!A({\bf x}\!\!-\!\!2{\bf r}))^2\rangle~'
\end{eqnarray}
These functions are roughly proportional to $r^{m-1}$ for a spectral slope $\propto k^{-m}$ if $m<5$ ($S\!F_2^{(3)}$) and $m<9$ ($S\!F_2^{(5)}$) respectively.   
Fig.~\ref{fig:3PSF}  reports the structure functions for the magnetic and the density fields measured in the local frame: this is defined by the parallel to the local mean direction ${\bf B_L}$, which is the average field $\langle {\bf B}\rangle$, seen in the interval $({\bf x},{\bf x}+{\bf r})$ and by the fluctuations ${\bf \delta B}$: $\hat{l}={\bf B_L}/{B_L}$ (dash-dotted lines) is the unit vector parallel to the local mean direction, $\hat{\lambda}$ (solid) is the unit vector normal to the both $\hat{l}$  and to  ${\bf \delta B}$, while $\hat{\xi}=\hat{l}\times{\hat{\lambda}}$ (dashed) completes the reference frame. As expected the 2nd-order structure functions are more energetic in the perpendicular component:  for separations smaller than about $d_{\rm i}$ they follow approximatively a power-law whose index $m-1$  correspond to a spectral index $m=2.8$, not far to what observed in Fourier spectra.  The important result here is however related to the behaviour of the direction parallel to the local field, which follows the same power-law as the perpendicular counterpart, shifted roughly by a factor of five to larger scales. The results is quite robust being almost the same for both the 3-point and 5-point SFs and confirm that, in this simulation where a relatively strong mean field is present, global spectra reproduce the spectral anisotropy at local scales. It should be noticed also, that such result is consistent with \citet{Arzamasskiy_al_2019} in which a spectral anisotropy scaling as $k_\parallel\propto k_\perp$ with respect to the local mean field was observed in analogous hybrid kinetic simulations by using a different technique based on Fourier-space filtering.

The SFs here reported are computed using the fields recovered in the real space once the filtering procedure has been applied in Fourier space: In the same figure (dotted-lines) are also reported the SFs computed without the filtering to show the effect of the noise. Although strongly localized in Fourier space, i.~e. at small scales, the noise impacts significantly the SFs over a much larger interval in the separation scales and it is stronger as the statistic used for the SFs increases. The reason lies in the noise of  the high-$k$ modes which introduces in each grid point $i$ a fluctuations $\pm \Delta f $ in the computation of the $S\!F_2 (f)$ of the quantity $f$: with $N$ the number of sampling point the quantity $f^2$ is affected by a systematic noise $\propto N^2$; the  SFs, which are $\propto \sum_i f^2/N$, are thus affected by a systematic noise at all scales $\propto N$. We have verified that it is indeed the case in this simulation. 

\begin{figure}[t]
\begin{center}
\includegraphics[width=0.78\linewidth]{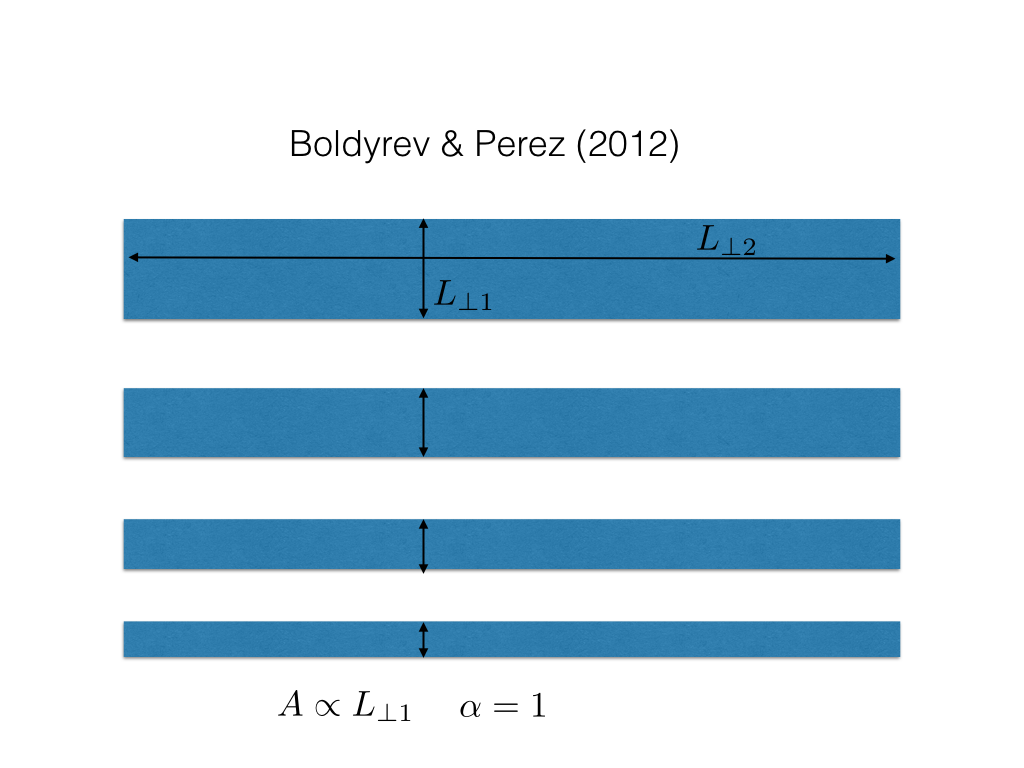}
\includegraphics[width=0.78\linewidth]{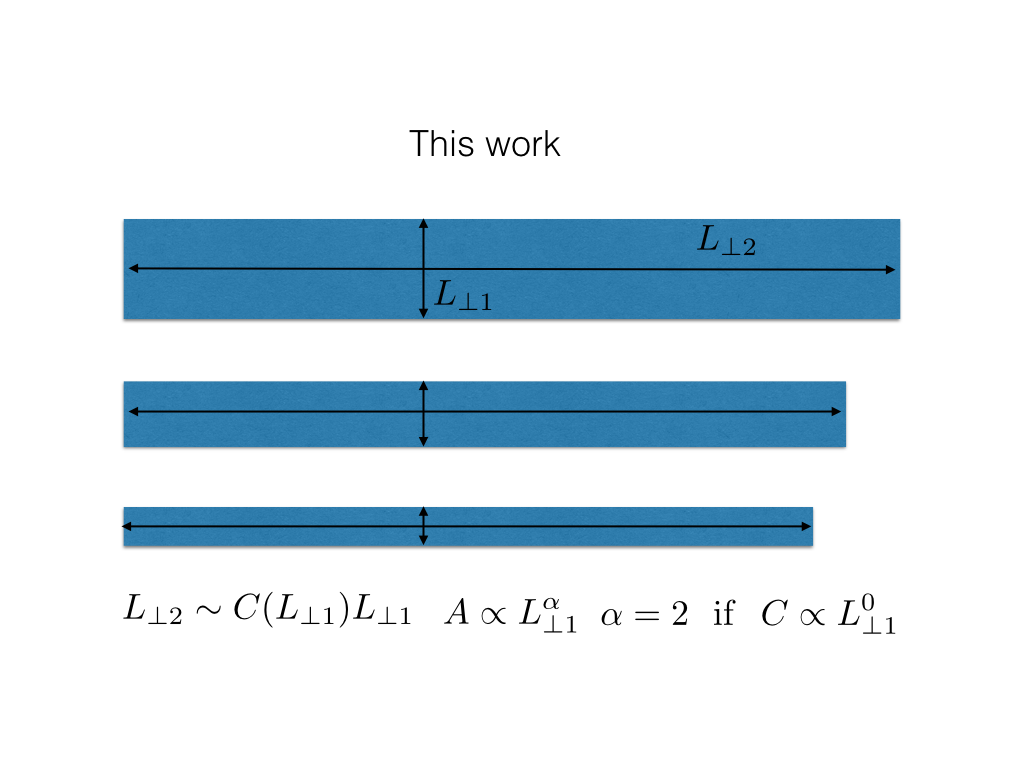}
\end{center}
\caption{Sketch of the filling factor scaling in \citet{Boldyrev_Perez_2012} (top) and the generalised scaling proposed in this work (bottom).}
\label{fig:CS12}
\end{figure}

\section{Discussion}
Both the global analysis with Fourier spectra and the one based on 2nd-order structure functions in the local magnetic field frame show that the scaling of the spectral anisotropy  is significantly less than what expected from current theories of turbulence at sub-ion scales. In particular we observe a scale independent anisotropy once the ion scales are reached. 
It is worth underlying that although the growth of the spectral anisotropy is reduced, the 2D Fourier spectrum of the magnetic field shows that the non-linear energy transfer happens in the perpendicular direction and is only advected along the parallel direction, thus suggesting that, like in the standard model of the critical balance \citep{Goldreich_Sridhar_1995} and in RMHD models of turbulence \citep{Matthaeus_al_1998,Oughton_al_1998}, the non-linear energy transfer time-scale is bounded by the decorrelation time which turns out to be that of the linear wave-propagation time.

A  possible interpretation of the results follows and extends the arguments of \citet{Boldyrev_Perez_2012} based on the presence of intermittent features \citep{Frisch_1995} localized in a 2D space normal to the local mean field. 
Following that argument, let us assume that at the perpendicular scale  $\lambda\propto k_{\perp}^{-1}$, the energy density fills only a fraction $\lambda^\alpha$ of the space. In this case the magnetic energy density scales as $E_\lambda\propto b^2_\lambda\lambda^{\alpha}$. Then, assuming that the energy transfer time at sub-ion scales is the non linear time $\tau_{\rm NL}\propto \lambda^2/b_\lambda$ \citep[e.~g.][]{Biskamp_al_1996}, and requiring that the energy flux is conserved through the scales, $E_\lambda/\tau_{\rm NL}\propto {\rm const}$, we have
\begin{equation}
b_\lambda \propto \lambda^{(2-\alpha)/3} ~{\rm ,}%\propto k_\perp^{-\frac{1}{3}(2-\alpha)}
\end{equation}
from which 
\begin{equation}
E_\lambda \propto \lambda^\alpha b_\lambda^2 \propto \lambda^{(4+\alpha)/3}~{\rm .}
\end{equation}
Therefore the power density spectrum $P_B$ goes as
\begin{equation}
P_B \propto  E(k_\perp)/k_\perp \propto k_\perp^{-\frac{1}{3}\left(7+\alpha\right)}~{\rm .}
\label{eq:alpha}
\end{equation}
The requirement that the non linear time is bounded by the decorrelation time, assuming the latter given by dispersive modes \citep[e.~g.,][]{Cho_Lazarian_2009}, requires for the parallel scale  $l\propto \lambda/b_\lambda$ or $k_\parallel \propto k_\perp b(k_\perp)$ from which the spectral anisotropy is 
\begin{equation}
k_{\parallel}\propto k_{\perp}^{\frac{1}{3}\left(\alpha +1 \right)}~{\rm .}
\label{eq:alpha3}
\end{equation}
In terms of the parallel scale the energy density goes as
\begin{equation}
E_l\propto l^{\frac{4+\alpha}{1+\alpha}}
\end{equation}
and the power density spectrum $P_B$ goes as
\begin{equation}
P_B \propto  E(k_\parallel)/k_\parallel \propto k_\parallel^{-\frac{5+2\alpha}{1+\alpha}}~{\rm ,}
\label{eq:alpha_par}
\end{equation}
For $\alpha=0$ one recovers the standard $P_B\sim k_\perp^{-7/3}$ and $k_\parallel\propto k_\perp^{1/3}$, while taking the filling factor proportional to $\lambda$ ($\alpha=1$) \citet{Boldyrev_Perez_2012} found $P_B\sim k_\perp^{-8/3}$ and a reduced anisotropy $k_\parallel \propto k_\perp ^{2/3}$. 
Interestingly, assuming that the filling factor scales as $\lambda^2$ ($\alpha=2$), the power given by Eq.~(\ref{eq:alpha}) will go as $k_\perp^{-3}$,  $B(k_\perp)\propto~{\rm const}$  and, following Eq.~(\ref{eq:alpha3}) the anisotropy will become scale-independent, i~e. $k_\parallel \propto k_\perp$.

A possible physical interpretation of such a result is illustrated in Fig.~\ref{fig:CS12}: in the plane perpendicular to the mean field, a structure has typically two characteristic lengths $l_{\perp 1}$ and $l_{\perp 2}$. In the model of \citet{Boldyrev_Perez_2012} (top panel) only one characteristic length (say $l_{\perp 1}$) changes with the scale, so that the filling factor reduces linearly as $l_{1\perp}$. However if the two characteristic lengths are correlated (for example if a specific aspect ratio of these structures is set as a function of their size) both will change with decreasing scale and the associated filling factor will be then reduced by a factor $l_{1\perp}l_{2\perp}\propto l_{\perp 1}^\alpha$ with $\alpha > 1$. In the special case that the aspect ratio is the same irrespective of the scale, we have $l_{\perp 2}={\rm C}l_{\perp 1}$ and $\alpha=2$;  in this case the power-law spectrum will scale as $k_\perp^{-3}$ and the spectral anisotropy will go as $k_\parallel \propto k_\perp$ in very good agreement with our results. 
Note that assuming $\lambda^\alpha=\lambda\xi$ implies $\lambda\propto\xi^{1/(\alpha-1)}$ (valid only if $\alpha\ne 1$ when the second perpendicular dimension has a meaning) from which $E_\xi \propto\xi^{(4+\alpha)/3(\alpha-1)}$. If $\alpha=2$ also the spectrum of the second perpendicular component follows the same power-law, which is also in agreement with our simulation: however in this case it is not possible to discriminate from other models where both perpendicular components are equivalent. 

\section{Conclusion}
We have analized the spectral anisotropy properties of plasma turbulence in a high-resolution hybrid PIC 3D simulation. The setup of the simulation is similar to that of the simulation discussed in \cite{Franci_al_2018} except for the higher resolution at small scales: here the grid step has been chosen 4 times smaller to better cover the sub-ion scales where our analysis is focused. The global spectral properties at sub-ion scales are very similar to what already observed: a slope steeper than $-7/3$ and $-8/3$ predicted by wave and intermittency based models, consistent with what observed in the solar wind and planetary magnetospheres and with what already observed both in 2D and 3D simulations \citep{Franci_al_2015a,Franci_al_2015b,Franci_al_2016,Cerri_al_2016,Cerri_al_2017b,Franci_al_2018,Arzamasskiy_al_2019,Papini_al_2019a}.

The main aspect investigated in this work is that in the global frame, both the parallel and perpendicular spectra show the same scaling, $k^{-3}$, below the ion break.
This picture has been confirmed further by performing a local analysis, by means of multi-point SFs, suggesting that in the simulation the aspect ratio of the turbulent fluctuations at sub-ion scales is maintained approximately constant. We have then proposed a model, based on a generalization of the one by \cite{Boldyrev_Perez_2012}, in which the spectral slopes (and thus the spectral anisotropy) are related to the nature of the intermittent structures that populate the turbulence at small scales and how they fill the total volume.

This result is in favour of a model where intermittency driven by 2D structures in the plane perpendicular to the main field plays a dominant role in determining the spectral properties at kinetic scales; such structures should have two characteristic lengths, both reducing with decreasing the scale, in a way that satisfies a specific aspect ratio, which can also be scale-dependent.  One possibility is that the turbulent cascade is strongly mediated by magnetic reconnection events where the aspect ratio of the current sheets is the important parameter setting the efficiency of the reconnection process \citep[e~g.][]{Loureiro_al_2007,Bhattacharjee_al_2009,Pucci_Velli_2014,Landi_al_2015,Tenerani_al_2015b,DelSarto_al_2016a,PucciF_al_2017,DelSarto_Ottaviani_2017}.  It has been already shown in \citet{Franci_al_2017, Cerri_Califano_2017} and \citet{Papini_al_2019a} that the formation of a sub-ion range in 2D simulations is strongly correlated with the emergence of reconnecting events between larger scale MHD vortices. Moreover it has been shown that current sheet disruption can have consequences on the spectral properties both at MHD \citep{Loureiro_Boldyrev_2017a,Mallet_al_2017a,Boldyrev_Loureiro_2017,Dong_al_2018} and kinetic scales \citep{Mallet_al_2017b,Loureiro_Boldyrev_2017b}. Our present findings seem to support this scenario.

Solar wind observations \citep[e. g.][]{Alexandrova_al_2009,Bruno_al_2014} show that the magnetic (and density) fluctuations converge toward a $-2.8$ scaling at smaller scales, a behaviour also well captured by high resolution 2D simulations \citep{Franci_al_2015a,Franci_al_2015b,Franci_al_2016}. In the framework of the model presented in this work, this corresponds to $\alpha = 1.4$, implying that the aspect ratio of the structure is not self-similar but depends on the scales. In this case the spectral anisotropy in the sub-ion range it is expected to moderately continue to increase as $k_\parallel \propto k_\perp^{0.8}$ and the parallel spectrum would follow a $-3.25$ power law. 

Although measurements of the spectral anisotropy are possible in the solar wind and have been performed in the inertial range \citep{Horbury_al_2008, Podesta_2009, Wicks_al_2011}, this aspect has been more scarcely addressed at kinetic scales partly because of observational constraints.
Spectral anisotropy in the sub-ion range has been discussed by \citet{Chen_al_2010b} using Cluster measurements and using two-points 2nd-order structure functions.  They found that the perpendicular spectrum was steeper than the theoretical $-7/3$, but that the spectral index in the parallel direction was less steep than the one expected by standard wave-mediated turbulence and close to $-3$. We note that these values are consistent with the predictions from our model discussed above.
However, as also discussed by the authors, one should be careful in interpreting their result, since when using 2-point 2nd-order SFs the resulting spectral index is limited to $-3$. Indeed, as discussed in the introduction, a slope steeper than $-7/3$  in the perpendicular spectrum, combined with the critical balance condition and without intermittency, would result in very steep parallel spectrum and large spectral anisotropies at small scales. For example a $-8/3$ scaling will give $k_\parallel \propto k_\perp^{1/6}$ and a parallel spectrum $\propto k_\parallel^{-11}$. In the near future,  to discriminate between such models will require multi-point, multi-scale analysis space missions, as it has been proposed in the Plasma 2020 Decadal Survey \citep{Matthaeus_al_2019,Klein_al_2019,TenBarge_al_2019}.
%\added[id=SL2]{To fully understand the physical grounds on the non-linear energy transfer at sub-ion scales a local analysis based on more than 2 multi-point observations is mandatory. In this }

%which in t are problematic because spectra slopes approach or overtake $-3$: however \citet{Chen_al_2010b} found at sub-ion scales magnetic field spectra exhibit similar power-law in the parallel and perpendicular direction, in more qualitative agreement with our results than parallel spectra predicted by the wave mediated turbulence $k_\parallel^{-5}$ or $k_\parallel^{-7/2}$ 
%\footnote{It is to mention that $k_\parallel^{-5}$ is valid only if the critical balance condition is coupled with the $7/3$ scaling in the perpendicular direction. Steeper slopes, such as $8/3$, will give stronger anisotropies $k_\parallel \propto k_\perp^{1/6}$, and even steeper parallel spectra, $\propto k_\parallel^{-11}$ }. \footnote{The critical balance spectral anisotropy both at MHD and kinetic scales can be written as $k_{\parallel}\propto k_{\perp} b(k_\perp)$ with $b(k_\perp)$ the magnetic fluctuations amplitudes at the scale $k_\perp$; as the spectrum is steeper as the anisotropy increases, and, for a spectral slope $-2.8$ the critical balance predicts $k_\parallel\propto k_\perp^{0.1}$}.

\section{Acknowledgments} 
We thanks Tullio for stimulating discussions and Dr. Ku Fu from Tibet University to point out us useful improvements. 
The work has been  funded by Fondazione Cassa di Risparmio di Firenze through the projects {\em Giovani Ricercatori Protagonisti'} and the project {\em HYPERCRHEL}. LF was supported by the UK Science and Technology Facilities Council (STFC) grant ST/P000622/1.
PH acknowledges GACR grant 15-10057S. 
LM was supported by the Programme National PNST of CNRS/INSU co-funded by CNES.
Numerical simulations and data reduction was performed at the CINECA facilities under the programs {\em Accordo Quadro INAF-CINECA (2017-2019)} (grant C3A22a) and  ISCRA-B (grant HP10BP6XYP). 
This work was performed also using the Cambridge Service for Data Driven Discovery (CSD3), part of which is operated by the University of Cambridge Research Computing on behalf of the STFC DiRAC HPC Facility (www.dirac.ac.uk). 
%The DiRAC component of CSD3 was funded by BEIS capital funding via STFC capital grants ST/P002307/1 and ST/R002452/1 and STFC operations grant ST/R00689X/1. DiRAC is part of the National e-Infrastructure

%We acknowledge PRACE for awarding us access to
%resource Cartesius based in the Netherlands at SURFsara through the
%DECI-13 (Distributed European Computing Initiative) call (project
%HybTurb3D), and the CINECA award under the ISCRA initiative, for the
%availability of high performance computing resources and support
%(grants HP10C877C4 and HP10BUUOJM). .

%\bibliographystyle{apj-eid}
\bibliographystyle{apsrev4-1}
%\bibliography{references}
\bibliography{mybib}
%\nocite{*}

\end{document}